\RequirePackage{lineno}
\documentclass[aps,prl,twocolumn,showpacs,superscriptaddress,groupedaddress]{revtex4}  
\usepackage{graphicx}  % needed for figures
\usepackage[FIGTOPCAP]{subfigure}
\usepackage{stackengine}
\usepackage{dcolumn}   % needed for some tables
\usepackage{bm}        % for math
\usepackage{amssymb}   % for math
\usepackage{color}
\usepackage{ulem}

\begin{document}

\title{Interference effect between $\phi$ and $\Lambda(1520)$ 
production channels in
the $\gamma p \rightarrow K^+K^-p$ reaction near threshold}

\newcommand{\KUADDRESS}{Department of Physics, Korea University, Seoul 02841, Republic of Korea}
\newcommand{\OHIOADDRESS}{Department of Physics and Astronomy, Ohio University, Athens, Ohio 45701, USA}
\newcommand{\KYOTOADDRESS}{Department of Physics, Kyoto University, Kyoto 606-8502, Japan}
\newcommand{\KYOTOOCRADDRESS}{Office of Commnunity Relations, Faculty of Science, Kyoto University, Kyoto 606-8502, Japan}
\newcommand{\KONANADDRESS}{Department of Physics, Konan University, Kobe, Hyogo 658-8501, Japan}
\newcommand{\XFELADDRESS}{XFEL Project Head Office, RIKEN, Sayo, Hyogo 679-5143, Japan}
\newcommand{\SINICAADDRESS}{Institute of Physics, Academia Sinica, Taipei 11529, Taiwan }
\newcommand{\NSRRCADDRESS}{Light Source Division, National Synchrotron Radiation Research Center, Hsinchu, 30076, Taiwan }
\newcommand{\RCNPADDRESS}{Research Center for Nuclear Physics, Osaka University, Ibaraki, Osaka 567-0047, Japan}
\newcommand{\JASRIADDRESS}{Japan Synchrotron Radiation Research Institute, Sayo, Hyogo 679-5143, Japan}
\newcommand{\YAMAGATAADDRESS}{Department of Physics, Yamagata University, Yamagata 990-8560, Japan}
\newcommand{\NAGOYAADDRESS}{Kobayashi-Maskawa Institute, Nagoya University, Nagoya, Aichi 464-8602, Japan}
\newcommand{\TOHOKUADDRESS}{Research Center for Electron Photon Science, Tohoku University, Sendai, Miyagi 982-0826, Japan}
\newcommand{\CHIBAADDRESS}{Department of Physics, Chiba University, Chiba 263-8522, Japan}
\newcommand{\MIYAZAKIADDRESS}{Department of Applied Physics, Miyazaki University, Miyazaki 889-2192, Japan}
\newcommand{\OSAKAADDRESS}{Department of Physics, Osaka University, Toyonaka, Osaka 560-0043, Japan}
\newcommand{\TOKYOIADDRESS}{Department of Physics, Tokyo Institute of Technology, Tokyo 152-8551, Japan}
\newcommand{\GIFUADDRESS}{Department of Education, Gifu University, Gifu 501-1193, Japan}
\newcommand{\RIKENADDRESS}{RIKEN, The Institute of Physical and Chemical Research, Wako, Saitama 351-0198, Japan}
\newcommand{\PKNUADDRESS}{Department of Physics, Pukyong National University, Busan 48513, Republic of Korea}
\newcommand{\JINRADDRESS}{Joint Institute for Nuclear Research, Dubna, Moscow Region, 142281, Russia}
\newcommand{\MSUADDRESS}{National Superconducting Cyclotron Laboratory, Michigan State University, East Lansing, MI 48824, USA}
\newcommand{\WAKAYAMAADDRESS}{Wakayama Medical College, Wakayama, 641-8509, Japan}
\newcommand{\SASKAADDRESS}{Department of Physics and Engineering Physics, University of Saskatchewan, Saskatoon, SK S7N 5E2, Canada}
\newcommand{\KEKADDRESS}{High Energy Accelerator Organization (KEK), Tsukuba, Ibaraki 305-0801, Japan}
\newcommand{\MINESOTAADDRESS}{School of Physics and Astronomy, 
University of Minnesota, Minneapolis, MN 55455, USA}
\newcommand{\PROTEINADDRESS}{Institute for Protein Research, 
Osaka University, Suita, Osaka 565-0871, Japan}
\newcommand{\AKITAADDRESS}{Akita Research Institute of Brain and 
Blood Vessels, Akita 010-0874, Japan}
\newcommand{\NDAADDRESS}{Department of Applied Physics, 
National Defense Academy, Yokosuka, Kanagawa 239-8686, Japan}
\newcommand{\KAOADDRESS}{Department of Physics, National Kaohsiung Normal University, Kaohsiung 824, Taiwan}
\newcommand{\CONNEADDRESS}{Department of Physics, University of Connecticut, Storrs, CT 06269-3046, USA}
\newcommand{\FUKUIADDRESS}{Proton Therapy Center, Fukui Prefectural Hospital, Fukui 910-8526, Japan}
\newcommand{\JAEAADDRESS}{Advanced Science Research Center, 
Japan Atomic Energy Agency, Tokai, Ibaraki 319-1195, Japan}
\newcommand{\CHUNGCHENGADDRESS}{Department of Physics, 
National Chung Cheng University, Chiayi 62102, Taiwan}
\newcommand{\KRISSADDRESS}{Korea Research Insititute of Standards 
and Science (KRISS), Daejeon 34113, Republic of Korea}

\author{S.~Y.~Ryu}\affiliation{\RCNPADDRESS}
\author{J.~K.~Ahn}\affiliation{\KUADDRESS}
\author{T.~Nakano}\affiliation{\RCNPADDRESS}
\author{D.~S.~Ahn}\affiliation{\RIKENADDRESS}
\author{S.~Ajimura}\affiliation{\RCNPADDRESS}
\author{H.~Akimune}\affiliation{\KONANADDRESS}
\author{Y.~Asano}\affiliation{\XFELADDRESS}
\author{W.~C.~Chang}\affiliation{\SINICAADDRESS}
\author{J.~Y.~Chen}\affiliation{\NSRRCADDRESS}
\author{S.~Dat$\acute{\rm{e}}$}\affiliation{\JASRIADDRESS}
\author{H.~Ejiri}\affiliation{\RCNPADDRESS}
\author{H.~Fujimura}\affiliation{\WAKAYAMAADDRESS}
\author{M.~Fujiwara}\affiliation{\RCNPADDRESS}
\author{S.~Fukui}\affiliation{\RCNPADDRESS}
\author{S.~Hasegawa}\affiliation{\RCNPADDRESS}
\author{K.~Hicks}\affiliation{\OHIOADDRESS}
\author{K.~Horie}\affiliation{\OSAKAADDRESS}
\author{T.~Hotta}\affiliation{\RCNPADDRESS}
\author{S.~H.~Hwang}\affiliation{\KRISSADDRESS}
\author{K.~Imai}\affiliation{\JAEAADDRESS}
\author{T.~Ishikawa}\affiliation{\TOHOKUADDRESS}
\author{T.~Iwata}\affiliation{\YAMAGATAADDRESS}
\author{Y.~Kato}\affiliation{\NAGOYAADDRESS}
\author{H.~Kawai}\affiliation{\CHIBAADDRESS}
\author{K.~Kino}\affiliation{\RCNPADDRESS}
\author{H.~Kohri}\affiliation{\RCNPADDRESS}
\author{Y.~Kon}\affiliation{\RCNPADDRESS}
\author{N.~Kumagai}\affiliation{\JASRIADDRESS}
\author{P.~J.~Lin}\affiliation{\SINICAADDRESS}
\author{Y.~Maeda}\affiliation{\FUKUIADDRESS}
\author{S.~Makino}\affiliation{\WAKAYAMAADDRESS}
\author{T.~Matsuda}\affiliation{\MIYAZAKIADDRESS}
\author{N.~Matsuoka}\affiliation{\RCNPADDRESS}
\author{T.~Mibe}\affiliation{\KEKADDRESS}
\author{M.~Miyabe}\affiliation{\TOHOKUADDRESS}
\author{M.~Miyachi}\affiliation{\TOKYOIADDRESS}
\author{Y.~Morino}\affiliation{\KEKADDRESS}
\author{N.~Muramatsu}\affiliation{\TOHOKUADDRESS}
\author{R.~Murayama}\affiliation{\OSAKAADDRESS}
\author{Y.~Nakatsugawa}\affiliation{\KEKADDRESS}
\author{S.~i.~Nam}\affiliation{\PKNUADDRESS}
\author{M.~Niiyama}\affiliation{\KYOTOADDRESS}
\author{M.~Nomachi}\affiliation{\RCNPADDRESS}
\author{Y.~Ohashi}\affiliation{\JASRIADDRESS}
\author{H.~Ohkuma}\affiliation{\JASRIADDRESS}
\author{T.~Ohta}\affiliation{\RCNPADDRESS}
\author{T.~Ooba}\affiliation{\CHIBAADDRESS}
\author{D.~S.~Oshuev}\affiliation{\SINICAADDRESS}
\author{J.~D.~Parker}\affiliation{\KYOTOADDRESS}
\author{C.~Rangacharyulu}\affiliation{\SASKAADDRESS}
\author{A.~Sakaguchi}\affiliation{\OSAKAADDRESS}
\author{T.~Sawada}\affiliation{\SINICAADDRESS}
\author{P.~M.~Shagin}\affiliation{\MINESOTAADDRESS}
\author{Y.~Shiino}\affiliation{\CHIBAADDRESS}
\author{H.~Shimizu}\affiliation{\TOHOKUADDRESS}
\author{E.~A.~Strokovsky}\affiliation{\JINRADDRESS}
\author{Y.~Sugaya}\affiliation{\RCNPADDRESS}
\author{M.~Sumihama}\affiliation{\GIFUADDRESS}
\author{A.~O.~Tokiyasu}\affiliation{\TOHOKUADDRESS}
\author{Y.~Toi}\affiliation{\MIYAZAKIADDRESS}
\author{H.~Toyokawa}\affiliation{\JASRIADDRESS}
\author{T.~Tsunemi}\affiliation{\KYOTOADDRESS}
\author{M.~Uchida}\affiliation{\TOKYOIADDRESS}
\author{M.~Ungaro}\affiliation{\CONNEADDRESS}
\author{A.~Wakai}\affiliation{\AKITAADDRESS}
\author{C.~W.~Wang}\affiliation{\SINICAADDRESS}
\author{S.~C.~Wang}\affiliation{\SINICAADDRESS}
\author{K.~Yonehara}\affiliation{\KONANADDRESS}
\author{T.~Yorita}\affiliation{\RCNPADDRESS}
\author{M.~Yoshimura}\affiliation{\PROTEINADDRESS}
\author{M.~Yosoi}\affiliation{\RCNPADDRESS}
\author{R.~G.~T.~Zegers}\affiliation{\MSUADDRESS}

\collaboration{The LEPS Collaboration}

\date{\today}

\begin{abstract}
The $\phi$-$\Lambda(1520)$ interference effect 
in the $\gamma p\to K^+K^-p$ reaction
has been measured for the first time in the energy range 
from 1.673 to 2.173 GeV.
The relative phases 
between $\phi$ and $\Lambda(1520)$ production amplitudes were obtained
in the kinematic region where the two resonances overlap.
The measurement results support strong constructive interference when 
$K^+K^-$ pairs are observed at forward angles, but
destructive interference for
proton emission at forward angles. 
Furthermore, the observed interference effect 
does not account for 
the $\sqrt{s}=2.1$ GeV bump structure 
in forward differential cross sections for $\phi$ photoproduction. 
This fact suggests possible
exotic structures such as a hidden-strangeness pentaquark state, 
a new Pomeron exchange 
and rescattering processes via other hyperon states.
\end{abstract}

\pacs{13.60.Rj, 13.88.+e, 24.70.+s, 25.20.Lj}
\maketitle

%\linenumbers

The $\phi$-meson production has the unique feature of gluon dynamics 
as a result of OZI suppression due to the dominant 
$s\bar{s}$ structure of the $\phi$ meson, which is predicted to proceed via 
the Pomeron trajectory with $J^{PC}=0^{++}$ 
\cite{titov, titov2, vmd, sibirtsev, laget, donnachie, landshoff}.
Cross sections for diffractive $\phi$ photoproduction are then predicted to 
increase smoothly with photon energy. 
However, a bump structure
at $\sqrt{s}=2.1$ GeV in forward differential cross sections 
was first 
reported by the LEPS collaboration \cite{mibe}.
Despite extensive experimental efforts 
devoted for the photoproduction of $\phi$ mesons near the threshold,
the nature of the bump structure 
has not yet been explained in detail \cite{seray,dey}.
Kiswandhi {\it et al.}~\cite{kiswandhi}
suggested that the bump structure is the result of
an excitation of missing nucleon resonances. 
However, the bump structure observed 
from CLAS appears only at forward angles;
thus, a conventional resonance interpretation seems less likely
\cite{dey}.
Very recently, the LHCb collaboration \cite{lhcb}
claimed to have observed two $J/\psi~p$ resonances 
referred to as hidden-charm pentaquark states 
($c\overline{c}uud$) from $\Lambda_b^0$ decays. In $\phi$ photoproduction, 
a hidden-strangeness pentaquark state could also be 
searched for as a candidate for the forward bump structure. 
Recent theoretical studies further relate this to a coupling between the $\phi p$ and 
$K^+\Lambda(1520)$ channels, because the bump structure occurs very close to 
the threshold of $\Lambda(1520)$ production \cite{ozaki, hyryu}. 
The $\phi$-$\Lambda(1520)$ interference could also 
account for the bump structure, but it has not 
yet been measured in $K^+K^-p$ photoproduction.
The interference may be either positive (constructive) or 
negative (destructive), 
depending on the relative phase between 
the production amplitudes of $\phi$ and $\Lambda(1520)$.

Here, we report the measurement of the 
forward differential cross sections 
for $\phi$ and $\Lambda(1520)$ photoproduction 
and the relative phase angles 
between their photoproduction amplitudes.
The importance of this analysis includes the event selection for 
$\gamma p\to K^+K^-p$, which
was based on a kinematic fit. Furthermore, the yields of 
$\phi$ and $\Lambda(1520)$
were obtained from a simultaneous fit of the $m_{K^+K^-}$ 
and $m_{K^-p}$ invariant masses with lineshapes
from a Monte-Carlo simulation.
This self-consistent analysis enables the investigation of a potential
interference effect between $\phi$ and $\Lambda(1520)$.
To our knowledge, no interference measurement 
for this reaction has previously been 
reported in the literature. 

The experiment was carried out using the LEPS detector at the SPring-8
facility in Japan. 
Linearly polarized photons with the energy from 
1.5 to 2.4 GeV was produced using a laser backscattering
technique \cite{mura} with UV lasers.
The photon beam was incident on a 15-cm liquid-hydrogen target, in which
$K^+$, $K^-$ and $p$ particles were produced and then passed through 
the LEPS spectrometer with the standard configuration \cite{nakano}. 

With a full data set of LH$_2$
runs, a new analysis on $\phi$-$\Lambda(1520)$ photoproduction was
performed using kinematic fits and simultaneous fits on the $K^+K^-$ 
and $K^-p$ mass spectra with Monte-Carlo lineshapes.
To identify candidate events, at least two of the $K^+$, $K^-$, 
and $p$ tracks were required to be reconstructed 
using standard particle identification methods. 

%%%%%%%%%%%%%%%%%%%%%%%%%%%%%%%%%%%%%%%%%%%
\begin{figure}[h]
\centering
\vspace{0.2cm}
\topinset{(a)}{
  \includegraphics[width=0.234\textwidth]{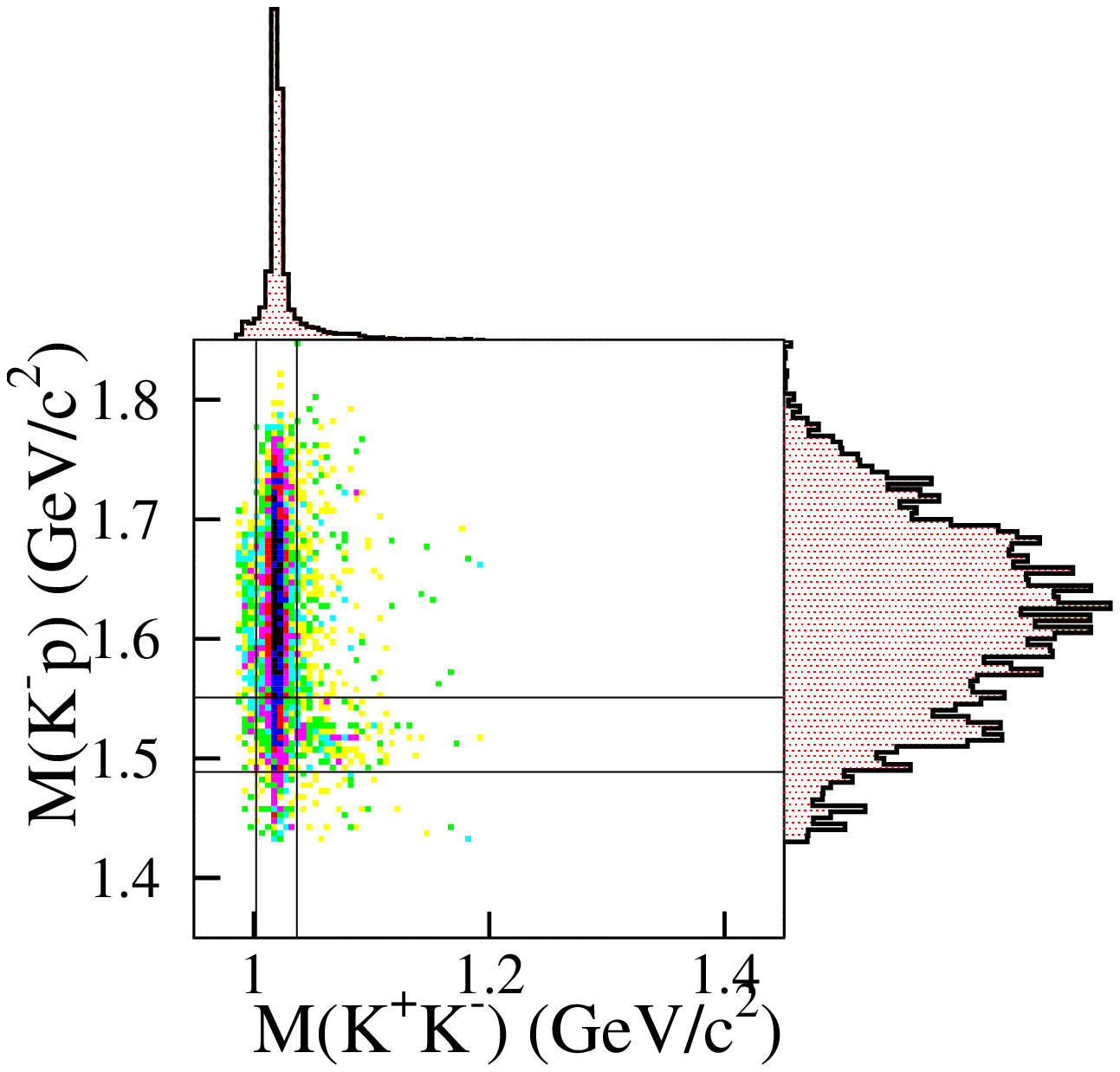}
  }{0.5cm}{1.0cm}
\hspace{-0.4cm}
\topinset{(b)}{
  \includegraphics[width=0.234\textwidth]{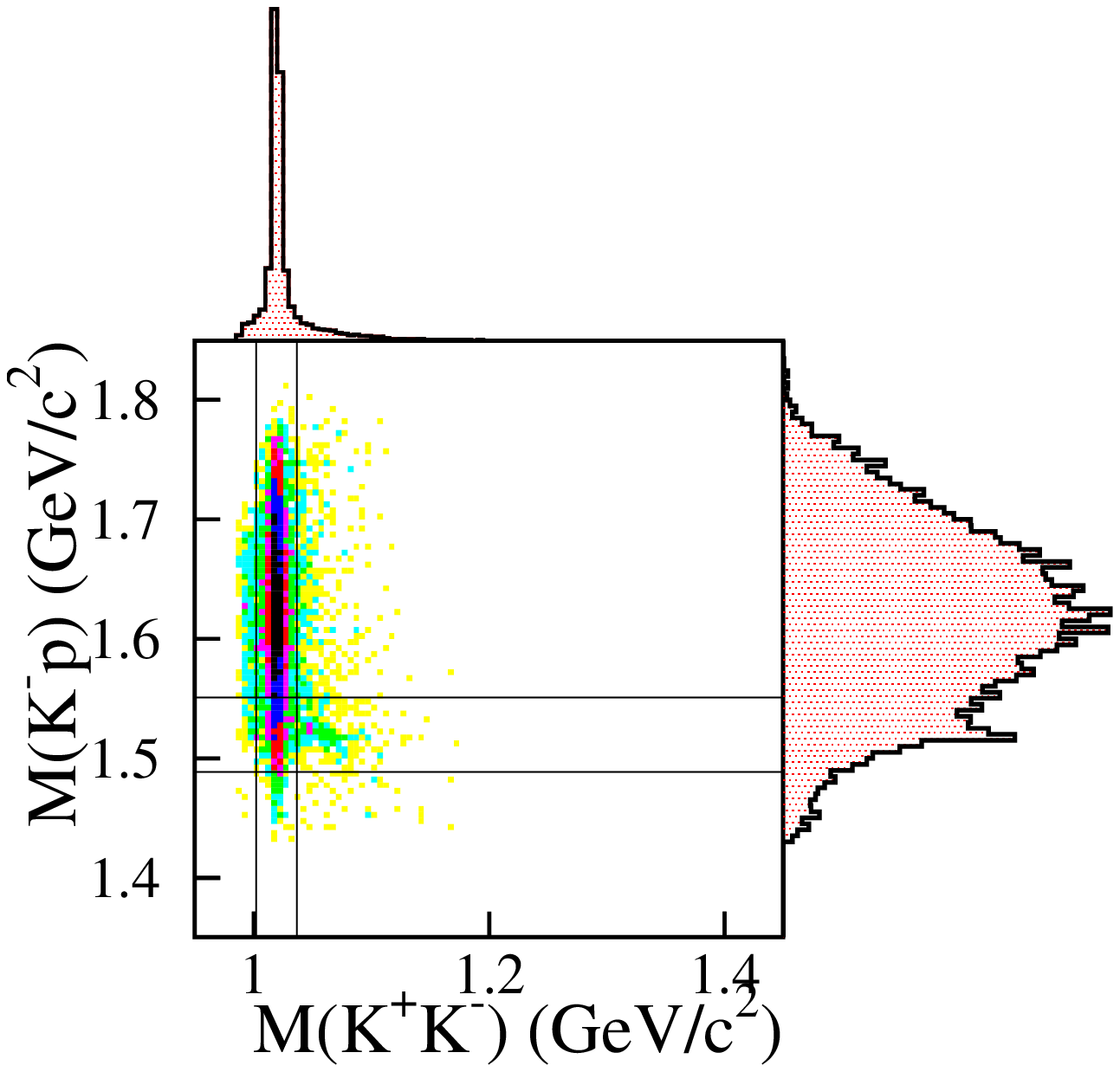}
  }{0.5cm}{1.0cm}
\\ \vspace{0.3cm}
\topinset{(c)}{
  \includegraphics[width=0.234\textwidth]{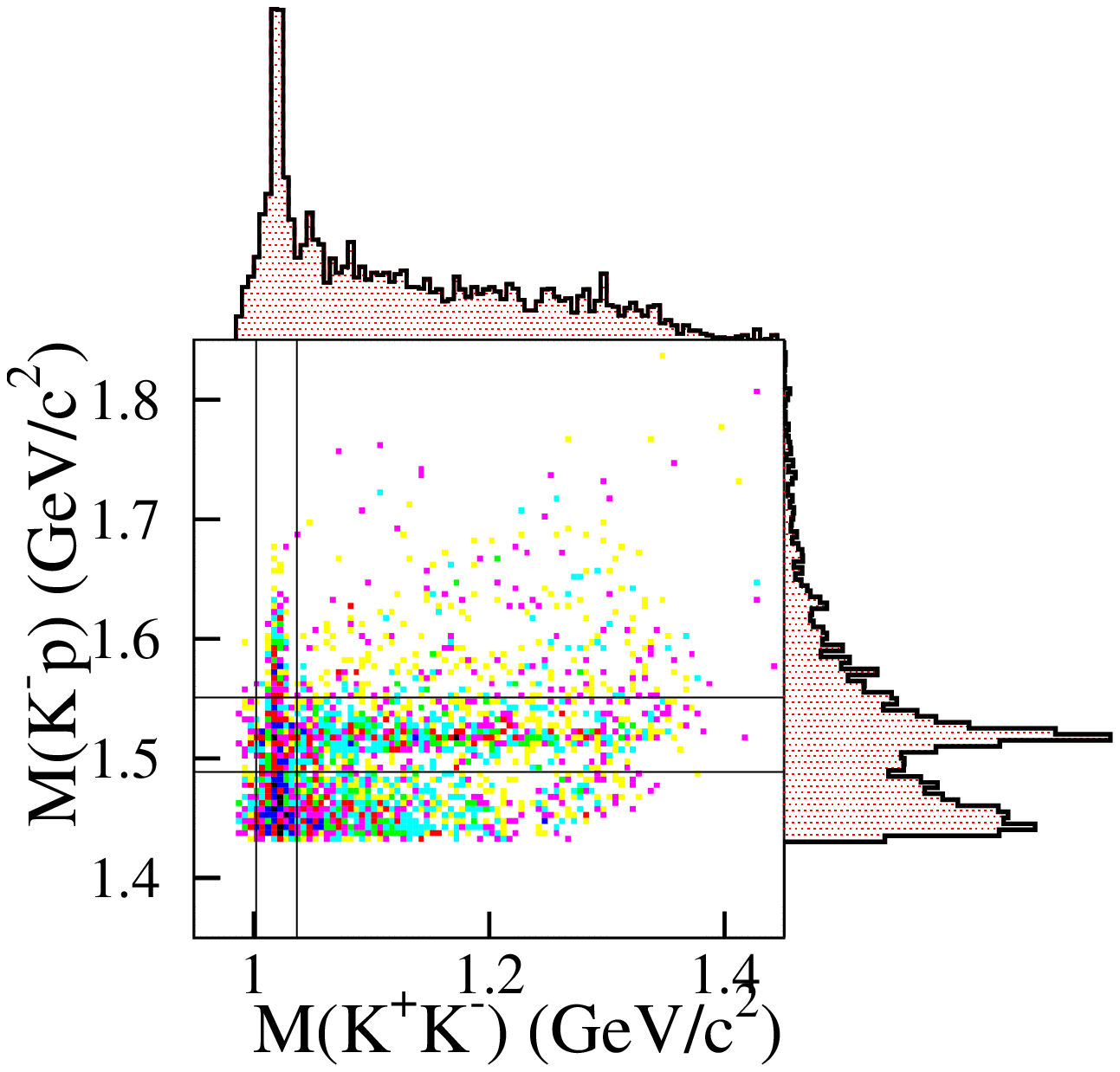}
  }{0.5cm}{1.0cm}
\hspace{-0.4cm}
\topinset{(d)}{
  \includegraphics[width=0.234\textwidth]{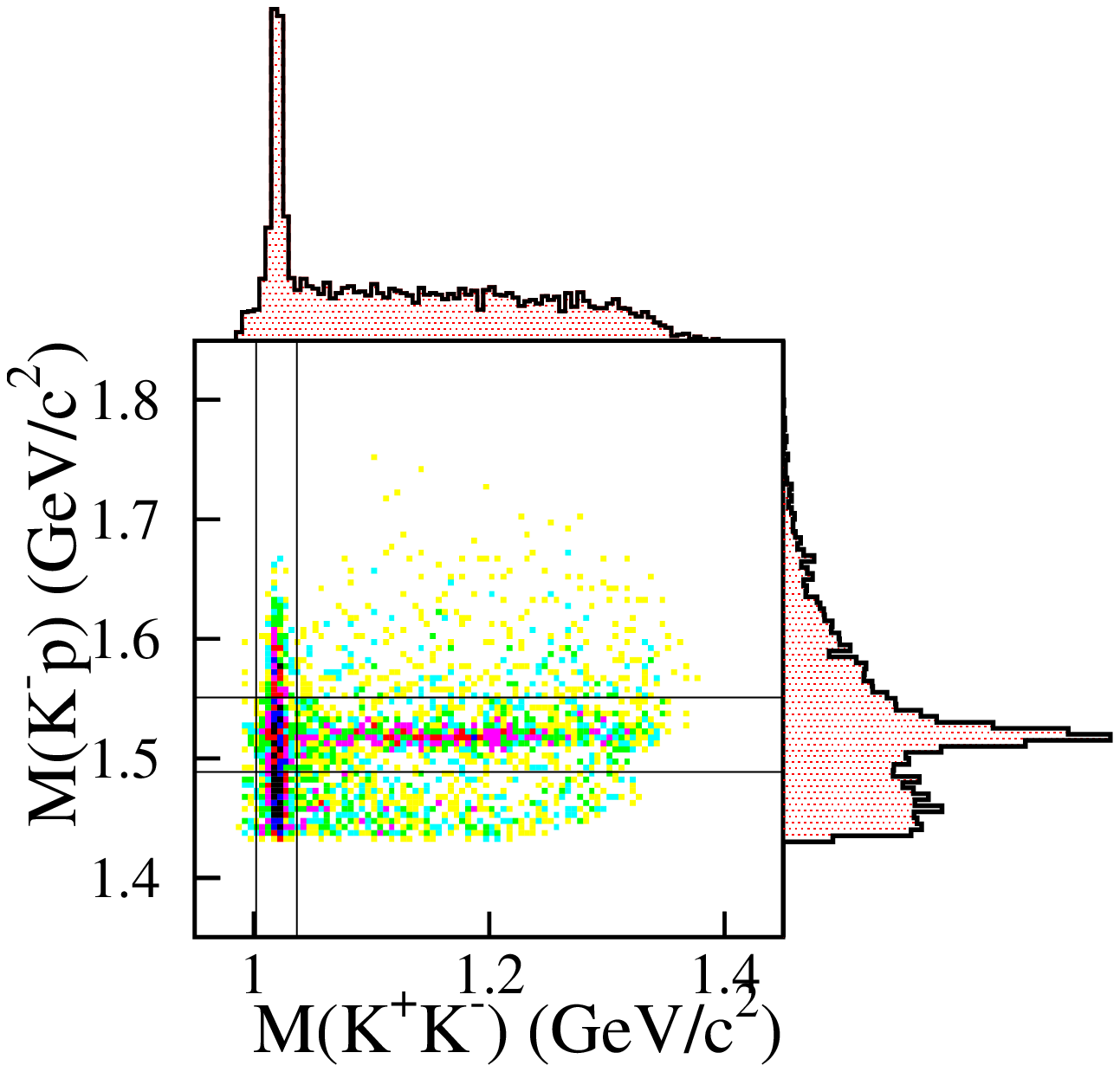}
  }{0.5cm}{1.0cm}
\\ \vspace{0.3cm}
\topinset{(e)}{
  \includegraphics[width=0.234\textwidth]{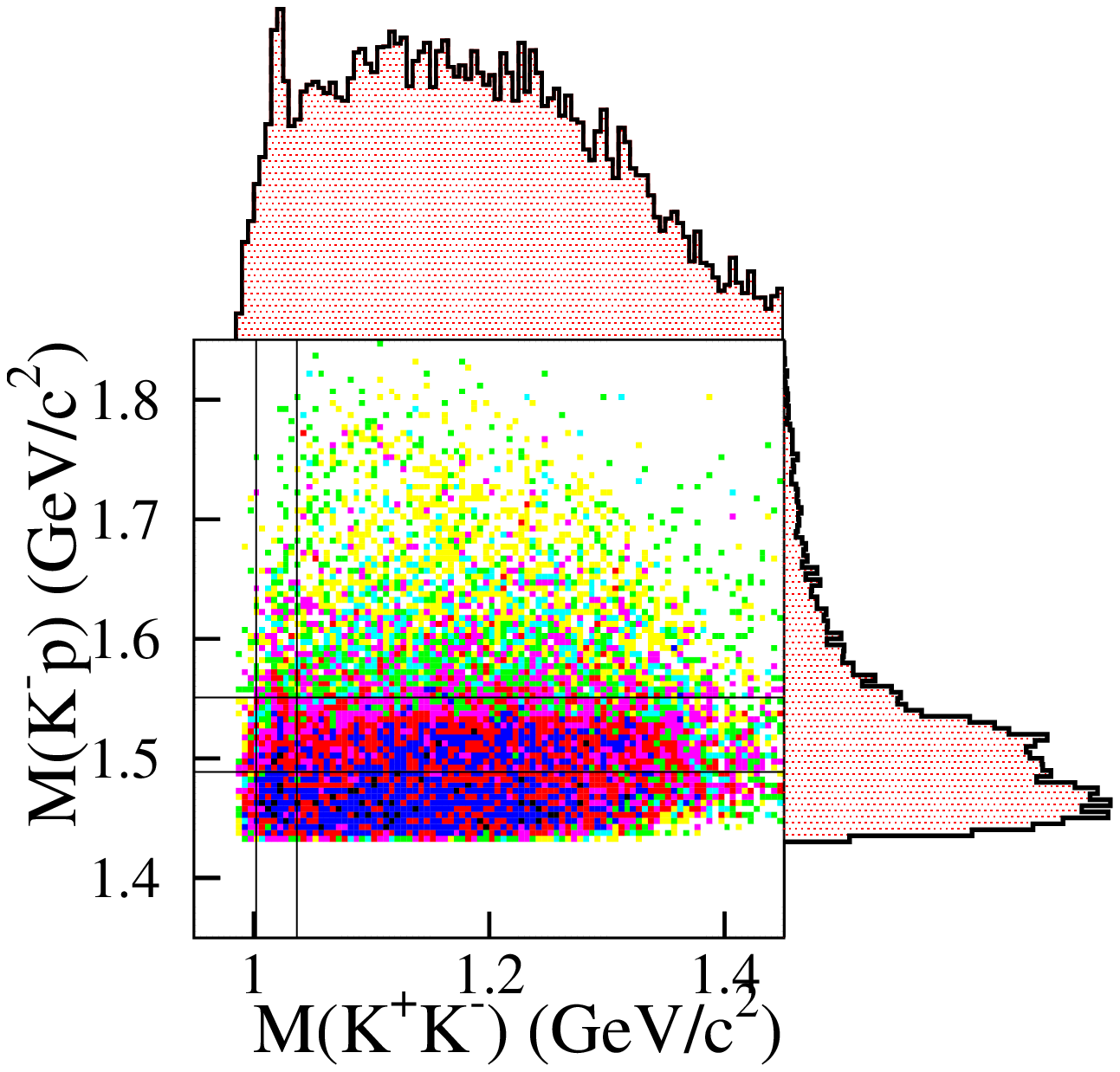}
  }{0.5cm}{1.0cm}
\hspace{-0.4cm}
\topinset{(f)}{
  \includegraphics[width=0.234\textwidth]{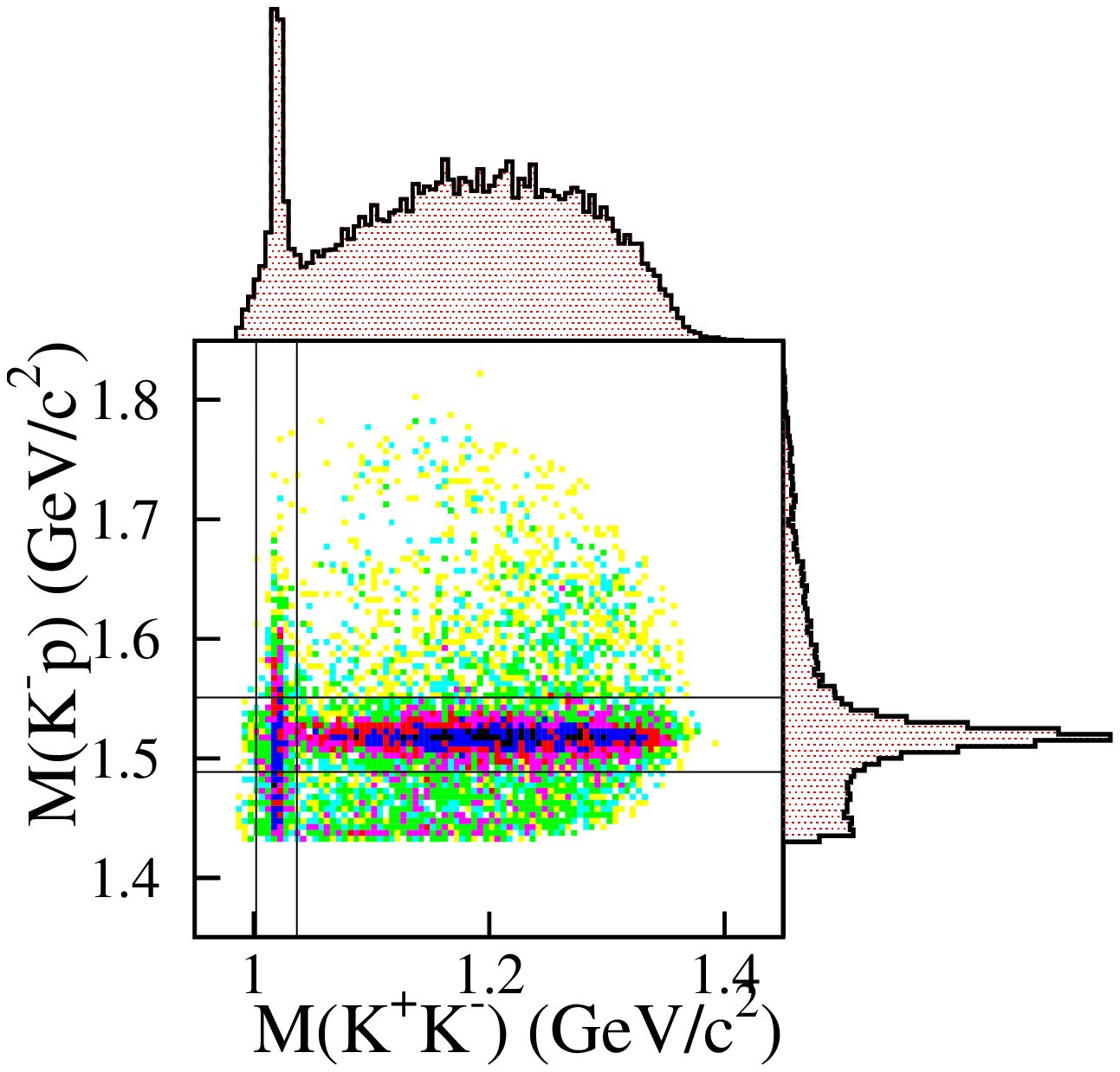}
  }{0.5cm}{1.0cm}
\caption{\label{fig1} Two-dimensional plot of 
the invariant mass of the $K^+K^-$ system versus that of 
the $K^-p$ system for forward $K^-K^+$ events (a) before and (b) after a
kinematic fit; (c), (d) and (e), (f) are same as (a), (b) but with 
forward $K^-p$ events and $K^+p$ events, respectively. 
Projections onto each mass axis are 
displayed at the upper and right sides.}
\end{figure}
%%%%%%%%%%%%%%%%%%%%%%%%%%%%%%%%%%%%%%%%%

Mass spectra, calculated from the measured four-vectors of the detected
$K^-$, $K^+$ and $p$, are shown in Fig. \ref{fig1}.
The solid lines represent the $\phi$ and $\Lambda(1520)$ mass bands, 
each corresponding to a $4~\Gamma_\phi$ window for $\phi$ production 
and a $2~\Gamma_{\Lambda^\ast}$ 
window for $\Lambda(1520)$ production, respectively, 
where $\Gamma_\phi=4.266$ MeV and 
$\Gamma_{\Lambda^\ast}=15.6$ MeV \cite{pdg}.
Forward particle pairs correspond to the pairs mostly in the range of
$\cos\theta^\ast >0.5$ in the production c.m. system.

The kinematic fit reconstructs three unmeasured parameters 
for a missing particle in the $K^-K^+p$ final state. 
The energy and momentum conservation laws provide four constraints. 
Consequently, we have an overdetermined system with four constraints 
and three unknowns.  
When the $\chi^2$-probability of the kinematic fit is required 
to be greater than 2\%, 
clear $\phi$ and $\Lambda(1520)$ bands 
are seen in the $M(K^+K^-)$ versus the $M(K^-p)$ (see Fig. \ref{fig1}(b), 
\ref{fig1}(d) and \ref{fig1}(f)). 
For forward $K^+p$ events (Fig. \ref{fig1}(e)), 
the background primarily represents
a $K^{\ast 0}\Sigma^+$ production channel with a small amount of 
$K^+\Lambda(1520)$ channel, followed by the
$\Lambda(1520)\to\Sigma^+\pi^-$ decay.
However, very little $K^\ast$ background remains
after a kinematic fit is applied (as shown by the histogram 
in Fig. \ref{fig1}(f)).

%%%%%%%%%%%%%%%%%%%%%%%%%%%%%%%%%%%%
\begin{figure}[h]
\centering
\hspace{-0.13cm}\includegraphics[width=0.44\textwidth]{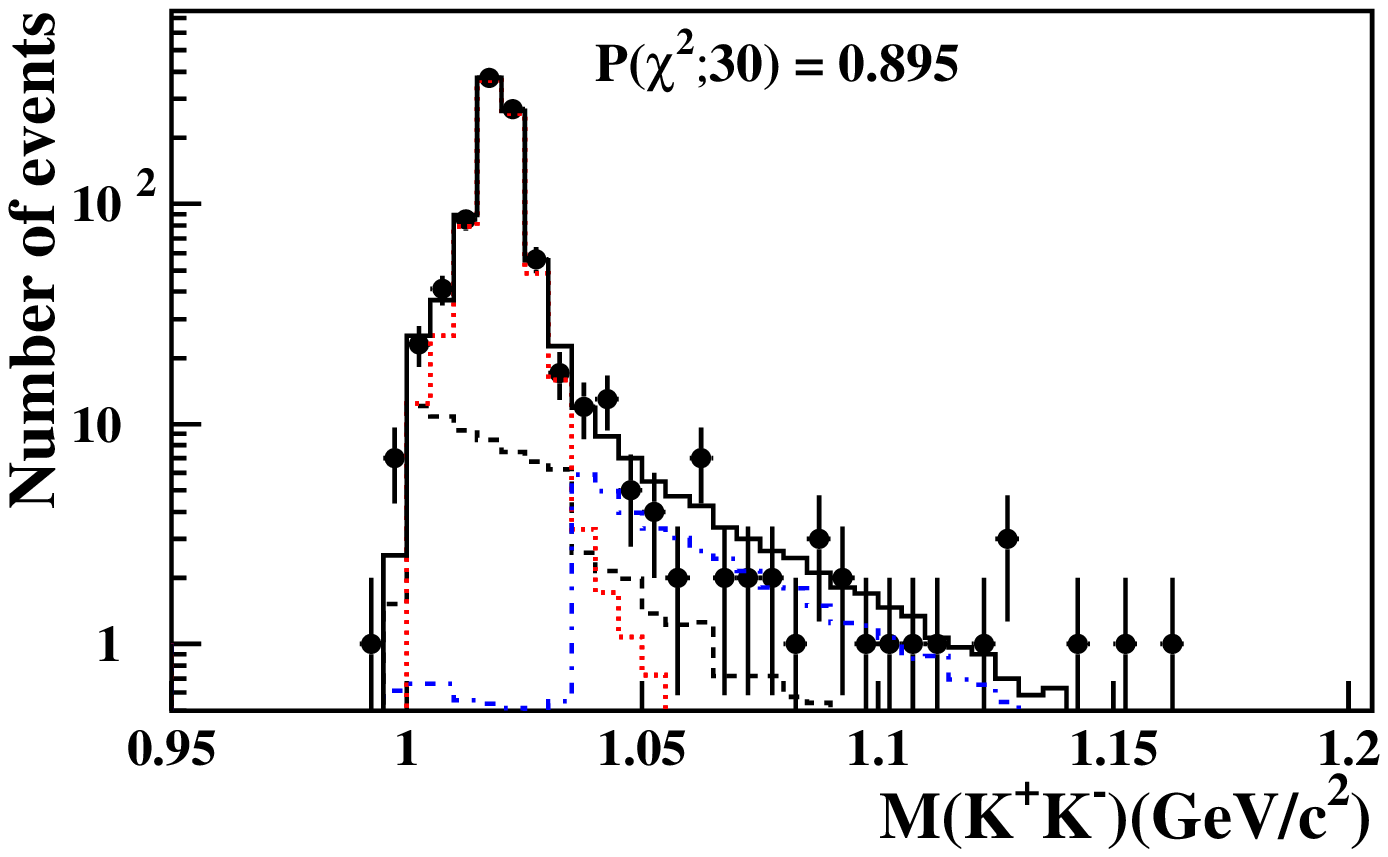}
\includegraphics[width=0.45\textwidth]{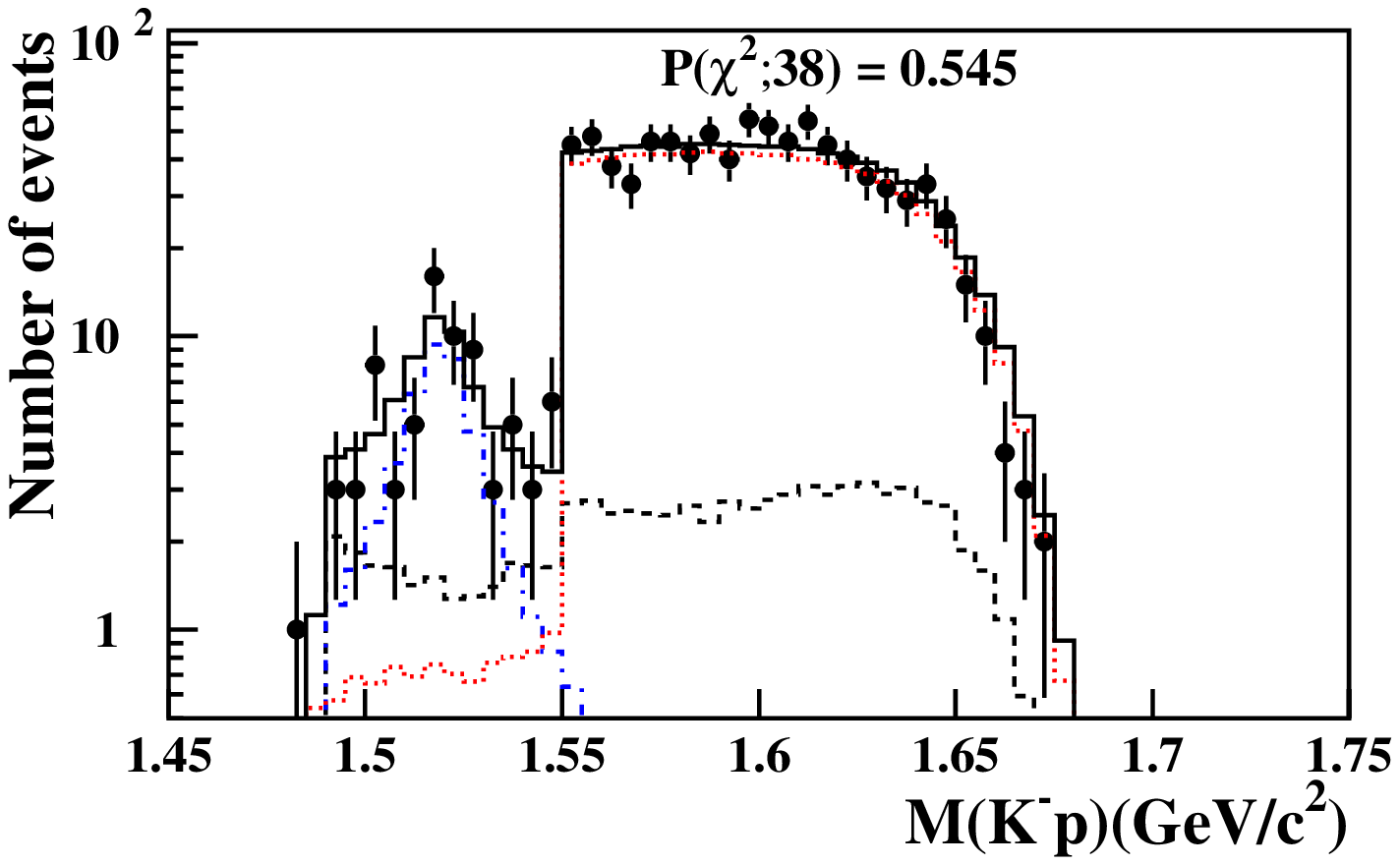}
\caption{\label{fig2}
The invariant mass spectra for (top) $K^+K^-$ and 
(bottom) $K^-p$ systems are displayed as closed circles for 
forward $K^+K^-$ events in the energy region from 1.973 to 2.073 GeV, respectively.
The best-fit lineshapes for $\phi$ are overlaid with dotted lines, 
while those for $\Lambda(1520)$ are represented as 
dot-dashed lines. Dashed lines 
represent the contributions of non-resonant $K^+K^-p$ production. }
\end{figure}
%%%%%%%%%%%%%%%%%%%%%%%%%%%%%%%%%%%%%%%%%

The measured $K^+K^-$ and $K^-p$ mass spectra for the selected 
$K^+K^-p$ events were fitted with the lineshapes of the simulated
processes, $\phi p$ \cite{wchang}, $\Lambda(1520)K^+$ \cite{chen}, 
and non-resonant $K^+K^-p$.
For events in which $K^+p$ is detected, these mass spectra are fitted 
with the three processes as well as 
$K^{\ast 0}\Sigma^+$ \cite{hwang} and 
$K^+(\Lambda(1520)\rightarrow\Sigma^+\pi^-)$.
The best-fit lineshapes for $\phi$, $\Lambda(1520)$ and non-resonant $K^+K^-p$ reproduce well 
both the $K^+K^-$ and the $K^-p$ mass spectra, 
as shown in Fig. \ref{fig2}.
The $\chi^2$ probability $P(\chi^2;{\it ndf})$
is quoted in each of the fitted $K^+K^-$ and $K^-p$ mass spectra, 
where {\it ndf} represents the number of degrees of freedom. 
The fits with Monte-Carlo lineshapes were based on the events
beyond the $\phi$-$\Lambda(1520)$ interference region 
in which the two resonances appear. The fit results
were then interpolated into the interference region, 
keeping the strengths of the Monte-Carlo lineshapes determined 
from the fit \cite{syryu}. 
This simultaneous fit with Monte-Carlo lineshapes is a 
self-consistent method to reproduce the measured $K^+K^-$ and $K^-p$
mass spectra, which pertains to the further study of 
interference effects.

%%%%%%%%%%%%%%%%%%%%%%%%%%%%%%%%%%%%
\begin{figure}[h]
\centering
%\subfigure[][]{
\includegraphics[width=0.231\textwidth]{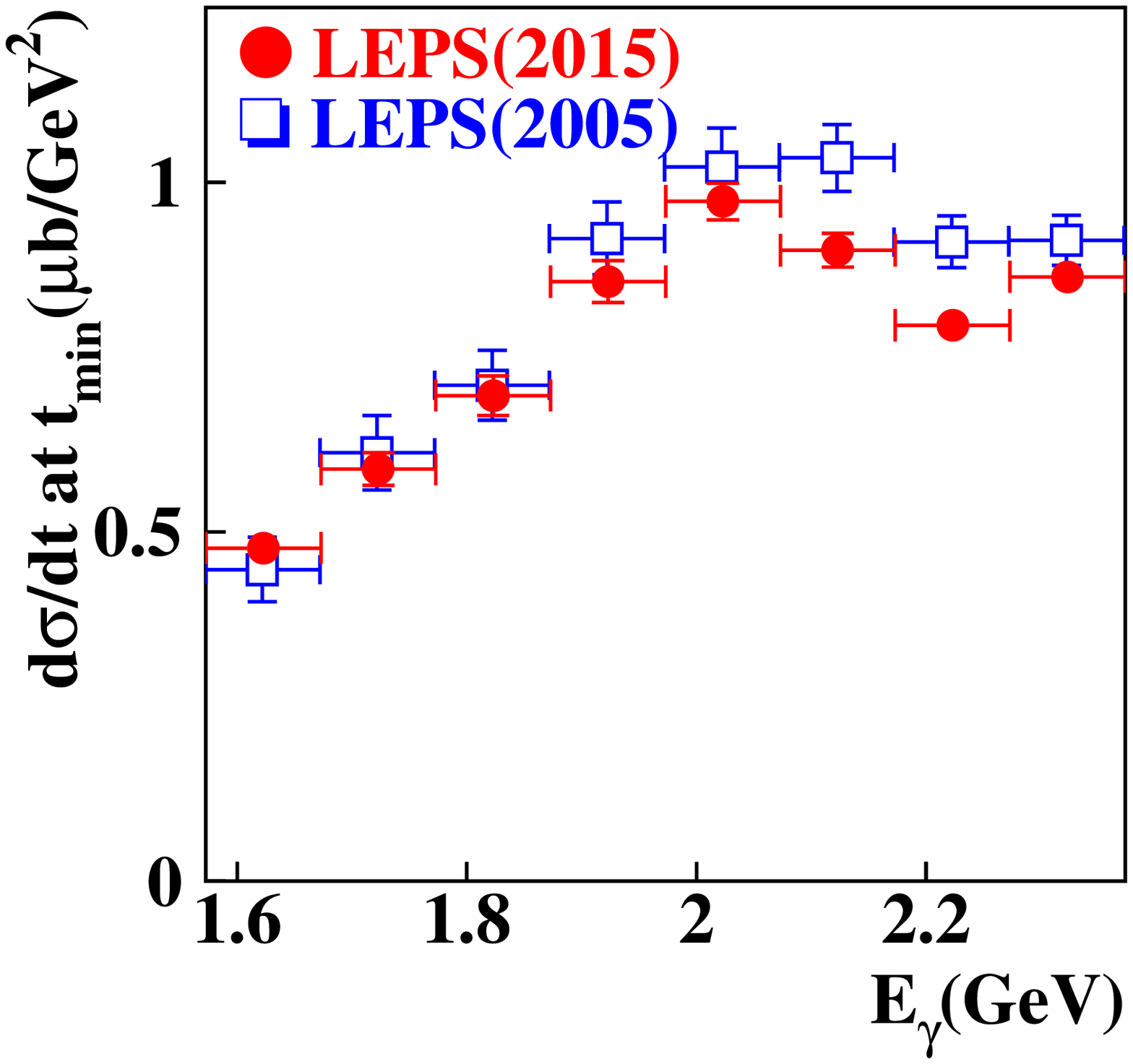}
%}
%\hspace{-0.35cm}
\hfill
%\subfigure[][]{
\includegraphics[width=0.24\textwidth]{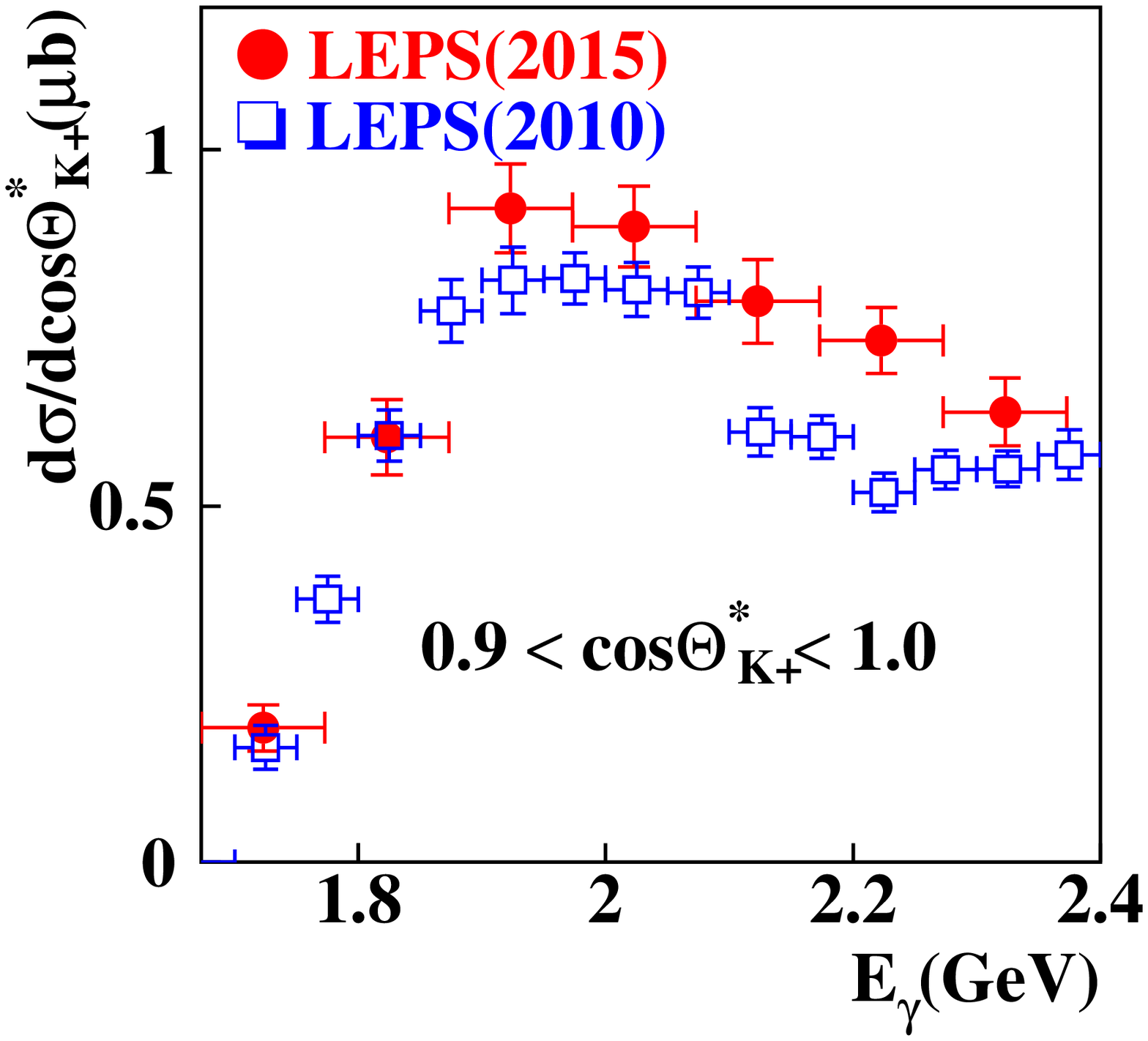}
%}
%\vspace{-0.4cm}
\caption{\label{fig3} 
Forward differential cross sections for (left) $\phi$
and (right) $\Lambda(1520)$ photoproduction. Open squares
represent cross-section results obtained without considering interference, 
while closed circles represent new measurement results.}
\end{figure}
%%%%%%%%%%%%%%%%%%%%%%%%%%%%%%%%%%%%%%%%%

Forward differential cross sections 
for $\phi$ and $\Lambda(1520)$ production channels
were measured using the best-fit results with Monte-Carlo lineshapes 
in the $\phi$ and $\Lambda(1520)$ mass bands except 
for the interference region.
The forward differential cross sections (d$\sigma/$d$t$ at $t=t_{\rm min}$)
for $\phi$ photoproduction are compared with previous results from 
LEPS \cite{mibe} near the threshold, as shown in Figure \ref{fig3}(left). 
Thus, 
we reconfirmed the existence of the bump structure 
around $E_\gamma = 2.0$ GeV. The structure appears persistently even with
different $\phi$-mass bands, 
different slope parameters, and the exclusion
of the interference region in which $\phi$ and $\Lambda(1520)$
mass bands overlap. The slope parameters of the $|t-t_{\rm min}|$ 
distributions decreased as the photon energy increased. 
The forward cross sections were obtained from the fit 
with linearly energy-dependent slope parameters 
(d$\sigma/$d$t=$d$\sigma/$d$t|_{t=t_{\rm min}}\exp(-b|t-t_{\rm min}|)$,
where $b=-(11.47-3.47~E)$ GeV$^{-2}$)
and $E$ is a dimensionless quantity
taken from the value of the photon energy in GeV.

Figure \ref{fig3}(right) shows differential cross sections
for $\Lambda(1520)$ photoproduction in the angular regions of
$0.9<\cos\theta_{K^+}^{\ast}<1.0$, which are compared with the 
previous LEPS results by Kohri {\it et al.} \cite{kohri}. 
While the previous analysis was based on events 
with a single $K^+$ track, this
measurement required at least two tracks among $K^-$, $K^+$, and $p$.
As a result, event statistics in this measurement 
at forward $K^+$ angles is smaller than that from previous
measurements. Though the statistics was low, both results are in good
agreement with each other, indicating the bump structure
near $E_\gamma=2$ GeV. 
Interestingly, the two cross-section results show the bump structure at
the same $E_\gamma$, which could indicate a strong correlation
between $\phi$ and $\Lambda(1520)$. 
However, the difference between the cross sections obtained 
with and without the interference region 
is not large enough to account for the bump structure. 
    
The differential cross sections for the $\gamma p\to K^+K^-p$ reaction
can be decomposed into
\begin{eqnarray}
\frac{d^2\sigma}{dm_{K^+K^-}dm_{K^-p}} 
\propto |\mathcal{M}_{\phi}+\mathcal{M}_{\Lambda(1520)}
+\mathcal{M}_{\rm nr}|^2, 
\end{eqnarray}
where $\mathcal{M}_\phi$ and
$\mathcal{M}_{\Lambda(1520)}$ are the complex amplitudes for
$\phi$ and $\Lambda(1520)$ production processes, respectively. 
$\mathcal{M}_{\rm nr}$ represents non-resonant $K^+K^-p$ production. 
Each complex amplitude includes individual amplitudes for 
all possible sub-processes, such as Pomeron-exchange and
pseudoscalar meson-exchange processes for $\phi$ photoproduction.
However, log-likelihood fits of the data in 
$\phi$ and $\Lambda(1520)$ bands 
excluding the $\phi$ and $\Lambda(1520)$ interference region 
($|\mathcal{M}_{\phi}+\mathcal{M}_{\rm nr}|^2$) 
with Monte-Carlo lineshapes
($|\mathcal{M}_{\phi}|^2+|\mathcal{M}_{\rm nr}|^2$) 
result in the $\chi^2$ probability
$P(\chi^2)>0.2$ in most cases. 
Moreover, the $S$-$P$ wave interference
in $\phi$ photoproduction is 
also known to be as small as the 1\% level \cite{fries}.
Therefore, we assume that
$|\mathcal{M}_{\phi}+\mathcal{M}_{\Lambda(1520)}
+\mathcal{M}_{\rm nr}|^2\approx  
|\mathcal{M}_{\phi}+\mathcal{M}_{\Lambda(1520)}|^2
+|\mathcal{M}_{\rm nr}|^2$, where 
the interference terms between 
$\mathcal{M}_{\rm nr}$ and two resonance amplitudes
are neglected. 
The contribution from the term $|\mathcal{M}_{\rm nr}|^2$ was then 
subtracted from the data.  

The differential cross sections for the $\gamma p\to K^+K^-p$ 
reaction via $\phi$ and $\Lambda(1520)$ resonances can be written as \cite{ejiri} 
%
%%%%%%%%%%%%%%%%%%%%%%%%%%%%%%%%%%%%%%%%%%%%%%%%%%
\begin{widetext}
\begin{eqnarray}
\frac{{\rm d}^2\sigma}{{\rm d}m_{K^{^+}K^{^-}}{\rm d}m_{K^{^-}p}}\Biggr|_{\phi,\Lambda(1520)}
~\propto~ |\mathcal{M}_{\phi}+\mathcal{M}_{\Lambda(1520)}|^2 
=
\Biggl|\frac{a ~e^{i\psi_a}}{m^2_\phi-m_{K^{^+}K^{^-}}^2+im_\phi\Gamma_\phi}
+\frac{b ~e^{i\psi_b}}{m^2_{\Lambda^\ast}-m_{K^{^-}p}^2
+im_{\Lambda^\ast}\Gamma_{\Lambda^\ast}}\Biggr|^2, 
\end{eqnarray}
\end{widetext}
%%%%%%%%%%%%%%%
where $a$ and $b$ denote the magnitudes of the Breit-Wigner amplitudes 
for $\phi$ and $\Lambda(1520)$, respectively. 
$\psi_a$ and $\psi_b$ represent phases 
for $\phi$ and $\Lambda(1520)$ production amplitudes, respectively.
Here, we integrate the differential cross sections over 
the $K^-p$ mass interval in
the $\phi$-$\Lambda(1520)$ interference region, 
assuming that the phase $\psi_b$ 
is constant in the interference region for each energy interval.
The integrated cross sections can then be given by
%%%%%%%%%%%%%%%%%%%%%%%%%%%%%%%%%%%%%%%%%%%%%%%%%%
\begin{eqnarray}
\frac{{\rm d}\sigma}{{\rm d}m}
\propto \Biggl|\frac{a~ e^{i\psi_a}}{m^2_\phi-m^2+im_\phi\Gamma_\phi}
+B(m)e^{i\psi_b}\Biggr|^2, 
\end{eqnarray}
%%%%%%%%%%%%%%%%%%%%%%%%%%%%%%%%%%%%%
where $m$ denotes $m_{K^{^+}K^{^-}}$. 
$|B(m)|^2$ corresponds to the Breit-Wigner lineshape
of $\Lambda(1520)$ projected onto the $K^+K^-$ mass axis
in the interference region. 
The interference term $I(m)$ 
between the two amplitude terms can be obtained as \cite{azimov}
%%%%%%%%%%%%%%%%%%%%%%%%%%%%%% 
\begin{eqnarray}
I(m)=2|aB(m)|
\frac{(m^2_\phi-m^2)\cos\psi
+\Gamma_\phi m_\phi\sin\psi}
{(m^2_\phi-m^2)^2+m^2_\phi\Gamma^2_\phi}, 
\label{eq:interference}
\end{eqnarray}
%%%%%%%%%%%%%%%
where $\psi=|\psi_a-\psi_b|$ is the relative phase between the phases,
$\psi_a$ and $\psi_b$.
%%%%%%%%%%%%%%%%%%%%%%%%%%%%%%%%%%
\begin{figure}[!htpb]
\centering
\includegraphics[width=0.48\textwidth]{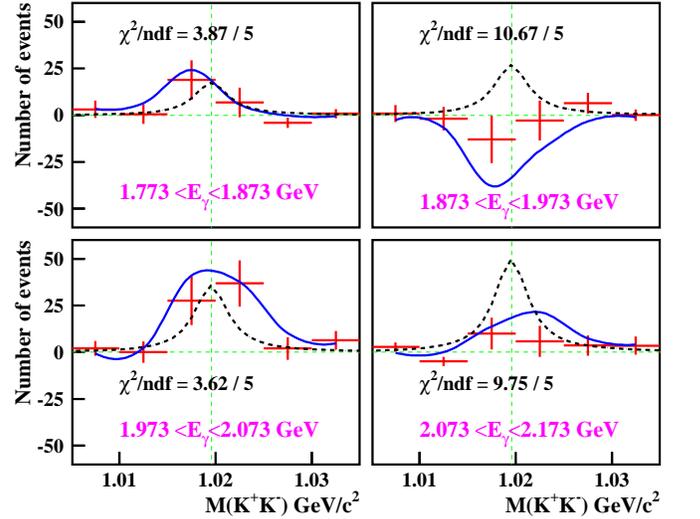}
\caption{\label{fig4} 
Measured event yields in the interference region are compared to 
the sum of $\phi$ and $\Lambda(1520)$ event yields for forward $K^+K^-$ events  
at four energy regions: from 1.773 GeV to 2.173 GeV with 0.1-GeV intervals.
The best-fit results for relative phase 
are overlaid with solid curves, while dashed lines indicate theoretical 
estimates assuming the maximum constructive 
$\phi$-$\Lambda(1520)$ interference with $\psi=\pi/2$.
}
\end{figure}
%%%%%%%%%%%%%%%%%%%%%%%%%%%%%%%%%%  

For the relative phase between the $\phi$ and $\Lambda(1520)$ 
amplitudes, we fitted data in the interference region 
with Eq.~\ref{eq:interference}. 
Here, the relative amplitudes of $a$ and $B(m)$ for each energy interval 
are fixed from a simultaneous fit utilizing Monte-Carlo lineshapes in
the $\phi$ and $\Lambda(1520)$ mass bands except for 
the interference region.
Consequently, only a single parameter, the relative phase $\psi$, 
exists in the fit. 
The best-fit results for relative phase are shown as 
solid curves in Fig. \ref{fig4}.  
To verify the reliability of this approach, the fit results are 
compared with theoretical estimates based on 
the effective Lagrangian approach \cite{sinam}, taking 
the two $\phi$ and $\Lambda(1520)$ production amplitudes into account. 
The reaction dynamics is represented
by the invariant amplitudes and form factors in this theoretical
approach. The phase of $\psi=\pi/2$ was chosen for simplicity.  
The theoretical estimates for the maximum constructive 
$\phi$-$\Lambda(1520)$ interference appear as dashed curves 
in Fig. \ref{fig4}, which are consistent with those predicted by 
Eq.~\ref{eq:interference}. 

%%%%%%%%%%%%%%%%%%%%%%%%%%%%%%%%%%%%%%%%%%%%%%%%%%%%%%%
\begin{figure}[h]
\centering
\includegraphics[width=0.238\textwidth]{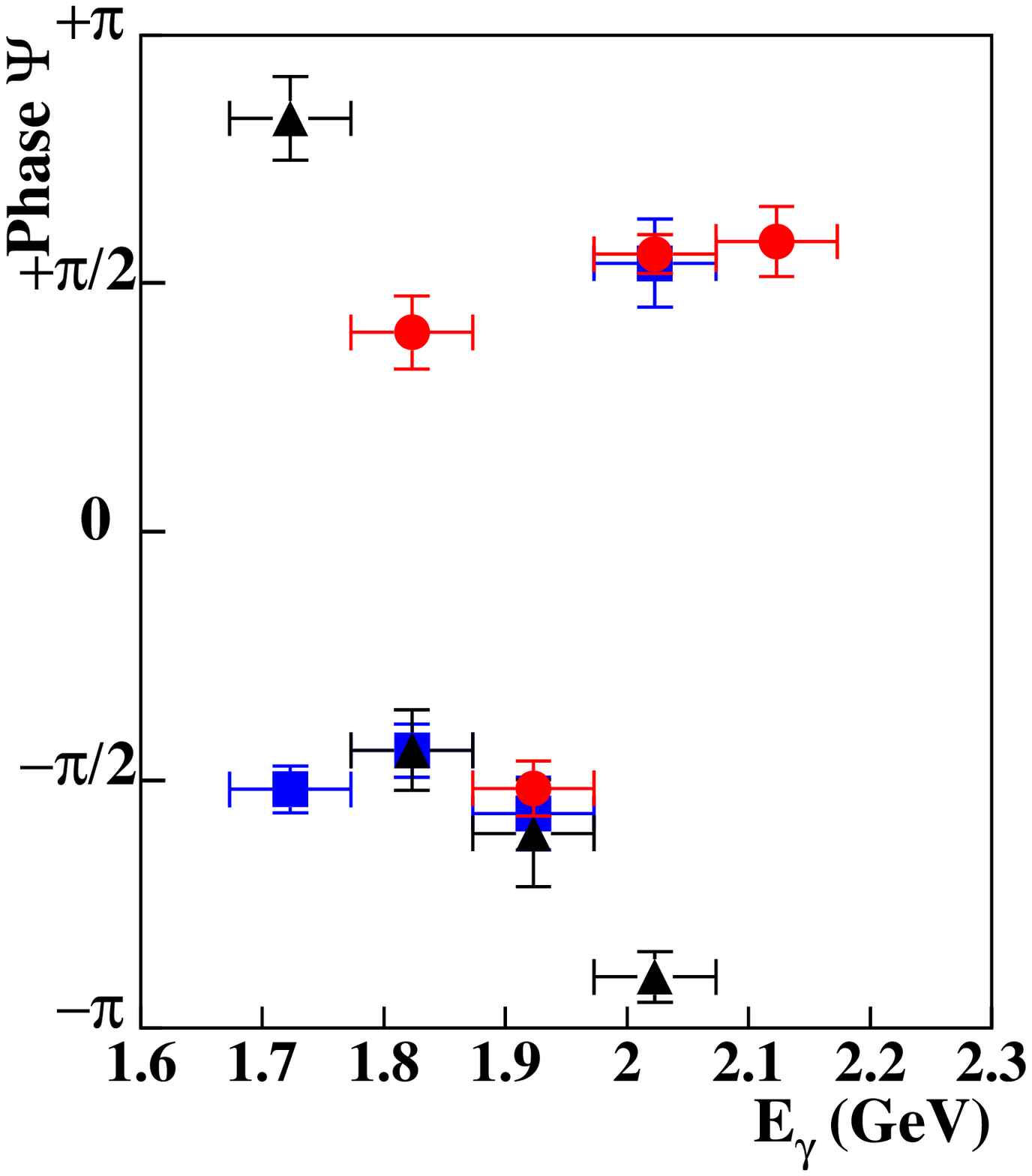}
\hfill
\includegraphics[width=0.238\textwidth]{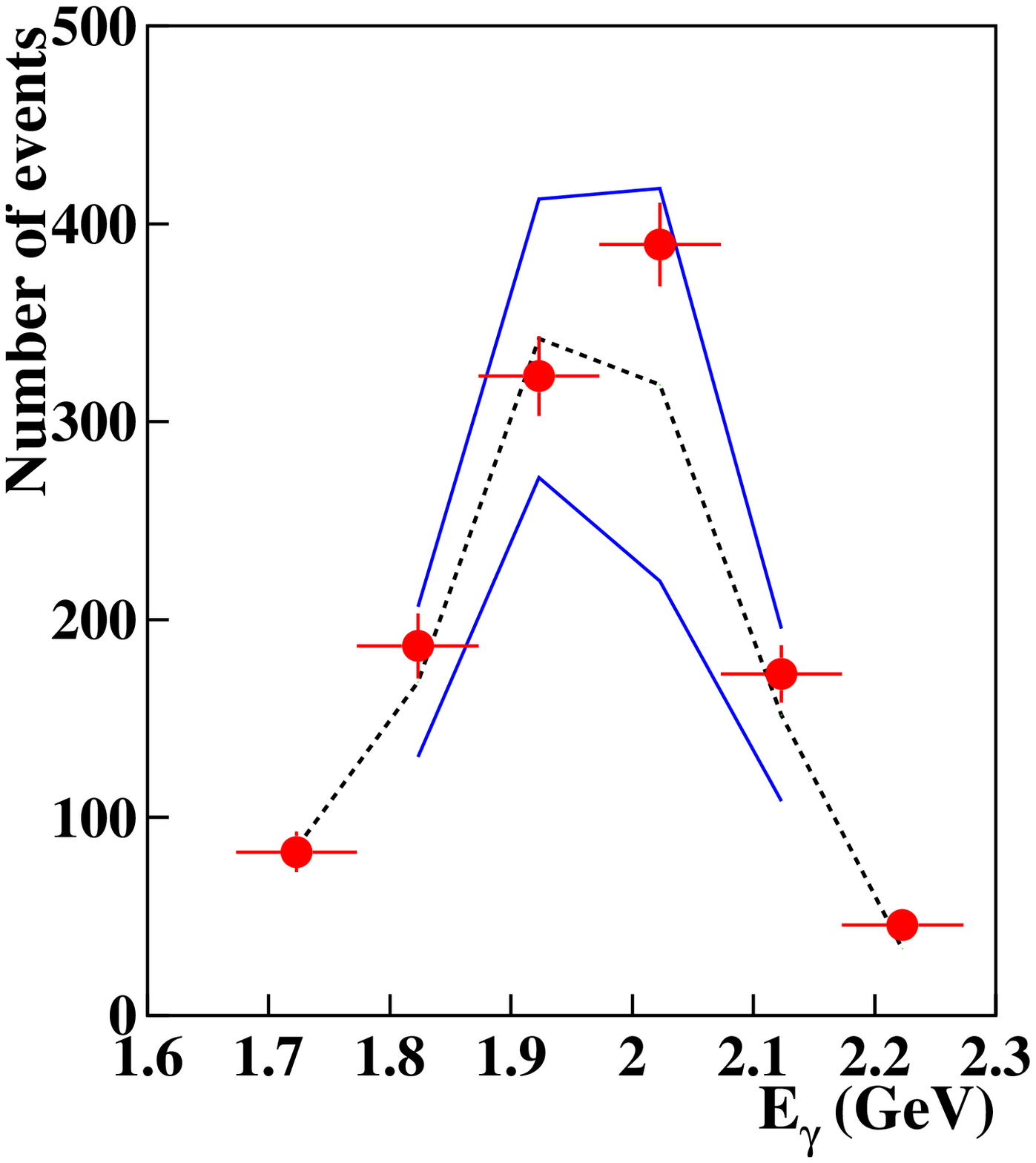}
\caption{\label{fig5} 
(Left) Phase angles for $K^+K^-$ (circles), 
$K^-p$ (squares), and $K^+p$ (triangles) events.   
(Right) Integrated yields (when $K^+K^-$ pairs are detected at forward
angles) in the interference region (circles)
compared to the predicted levels for the maximum and minimum bounds 
(solid lines). The dashed line indicates the predicted levels 
for no $\phi$-$\Lambda(1520)$ interference. 
}
\end{figure}
%%%%%%%%%%%%%%%%%%%%%%%%%%%%%%%%%%%%%%%%%%%%%%%%%%%%%%%

The fit results for relative phase are represented in Fig. \ref{fig5}(left).
The $\chi^2$ probability was required to exceed 0.1\%. 
For forward $K^-p$ and $K^+p$ events, the energy regions between
$1.673$ GeV and 2.073 GeV are explored.
The maximum constructive interference has $\psi=\pi/2$, 
while the maximum destructive interference is represented by $\psi=-\pi/2$. 
For $K^+K^-$ events detected in the forward directions, 
the resulting relative phases are in most cases constructive, while
those for forward $K^\pm p$ events are destructive. 

For forward $K^+K^-$ events in the energy region of 
$1.973< E_\gamma <2.073$ GeV,
the integrated event yield in the interference region approaches close 
to the maximum bound
for the $\phi$-$\Lambda$ interference, as shown 
in Fig. \ref{fig5}(right), which 
is consistent with the relative phase $\psi=1.69~\pm~ 0.12$ rad. 
Moreover, the relative phase flips its sign 
as a function of photon energy $E_\gamma$.  
For $K^-p$ events, the relative 
phase in the energy region of $1.973 <E_\gamma<2.073$ GeV 
firmly stays at a positive value, while 
in other energy regions it supports
destructive interference. Thus, it could be inferred that a change in 
interference patterns would occur when $K^-p$ moves to a forward angle. 
For the $K^+p$ events, only in the lowest-energy region 
the phase appears in the positive side, 
but it remains close to $\pi$, which corresponds to zero interference.

Different phases for the event mode (forward $K^+K^-$,
$K^-p$ and $K^+p$ events) may arise from differing kinematic
coverages for the photoproduction of $\phi$ and $\Lambda(1520)$. 
We could relate the phases near $\pi/2$ for forward $K^+K^-$ events to 
the interference between Pomeron exchange amplitude for 
$\phi$ and $K$-exchange amplitude for $\Lambda(1520)$ photoproduction.
For forward proton events ($K^-p$ and $K^+p$), unnatural-parity exchange
processes become important in $\phi$ photoproduction. 
However, it is worth noting that $\phi$-$\Lambda(1520)$ interference 
effect does not account for a 2.1-GeV bump structure in forward differential 
cross sections for $\phi$ photoproduction. This result is consistent with
a recent report from CLAS regarding the $\Lambda(1520)$ effect \cite{dey}.   
The energy dependence of the phase could indicate nontrivial rescattering contributions 
from other hyperon resonances.
The bump structure could then potentially be due to either rescattering
processes due to kinematic overlap in phase space or exotic structures involving 
a hidden-strangeness pentaquark state and the exchange of a new Pomeron.  
Alternatively, they could be due to a combination of both factors.

In summary, the photoproduction of the
$\gamma p \rightarrow K^+K^-p$ reaction was 
measured using the LEPS detector at energies from 1.57 to 2.40 GeV.
The $\phi$-$\Lambda(1520)$ interference measurement 
could be a good probe to study the origin of enhanced production
cross sections for $\phi$ and $\Lambda(1520)$ near $\sqrt{s}=2.1$ GeV.
In this Letter, 
we presented relative phases between $\phi$ and $\Lambda(1520)$
production amplitudes by using a two-dimensional mass fit 
with Monte-Carlo lineshapes.
We reconfirmed the bump structure in the analysis without
the $\phi$-$\Lambda(1520)$ interference region. 
On the other hand, we observed clear 
$\phi$-$\Lambda(1520)$ interference effects
in the energy range from 1.673 to 2.173 GeV. 
The data obtained in the present study 
provide the first-ever experimental evidence 
for the $\phi$-$\Lambda(1520)$ interference effect 
in $\phi$ photoproduction.
The relative phase results 
suggest strong constructive interference in most cases 
for $K^+K^-$ pairs observed at forward angles, 
while destructive interference results from the emission of
protons at forward angles. 
The nature of the bump structure could originate from 
interesting exotic structures such as a hidden-strangeness pentaquark state, 
a new Pomeron exchange and rescattering processes via other hyperon states.

%% acknowledgement  
The authors gratefully acknowledge the contributions of the staff of
the SPring-8 facility. Thanks are due to A. Hosaka for invaluable
theoretical discussions.
This research was supported in part by the US National Science
Foundation, the Ministry of Education, Science, Sports and Culture of
Japan, the National Science Council of the Republic of China, the
National Research Foundation of Korea, Korea University and Pukyong National University.
% (2012R1A2A1A01011926). 

\end{document}